\documentclass[10pt,twocolumn,a4paper,conference]{IEEEtran}
\IEEEoverridecommandlockouts
\usepackage[ruled,vlined]{algorithm2e}
\usepackage{cite,url}
\usepackage{amsmath,amssymb,amsfonts}
\usepackage[]{algorithm2e}
\usepackage{graphicx}
\usepackage{textcomp}
\usepackage{xcolor}
\usepackage{balance}
\usepackage{makecell}
\usepackage{ctable}
\usepackage{tabularx,booktabs}
\def\BibTeX{{\rm B\kern-.05em{\sc i\kern-.025em b}\kern-.08em
    T\kern-.1667em\lower.7ex\hbox{E}\kern-.125emX}}
\usepackage{multirow}
\usepackage{tabularx}
\usepackage{listings,xcolor}
\makeatletter
\def\thickhline{%
  \noalign{\ifnum0=`}\fi\hrule \@height \thickarrayrulewidth \futurelet
   \reserved@a\@xthickhline}
\def\@xthickhline{\ifx\reserved@a\thickhline
               \vskip\doublerulesep
               \vskip-\thickarrayrulewidth
             \fi
      \ifnum0=`{\fi}}
\makeatother

\newlength{\thickarrayrulewidth}
\newcommand{\refs}[1]{Section~\ref{#1}}

\newcommand{\reff}[1]{Fig.~\ref{#1}}
\newcommand{\reft}[1]{Table ~\ref{#1}}

\begin{document}
\title{ LiFi Technology Overview: taxonomy, and future directions}
\author{{Victor Monzon Baeza$^1$} and {Rafael Arellano Garcia$^2$}\\[2.5mm]
    {$^1$Interdisciplinary Centre for Security, Reliability and Trust (SnT), University of Luxembourg, Luxembourg}\\
    {$^2$Open University of Catalunya, Spain}\\
     {e-mail: {\tt victor.monzon@uni.lu}, {\tt rarellano@uoc.edu}}

\thanks{The authors are with $^1$University Carlos III of Madrid, Department of Signal Theory and 
 Communications e-mail: {\tt vmonzon@tsc.uc3m.es}}}
\date{}
\maketitle

\begin{abstract}
The looming electromagnetic spectrum crisis -due to the fact of the explosive growth in the increasing user data demand- has encouraged the emergence of new wireless technologies. This paper surveys the state-of-the-art leading and rapid developments in the current Light Fidelity (LiFi) technology. First, an overview is shown to help readers understand the potential of this technology. A comprehensive comparison with legacy wireless technologies is analyzed. 

We devise a taxonomy based on the main characteristics such as modulation techniques, applications, scenarios, network topologies, and architecture requirements, which enables a foundation for the research community. We highlight the powerful opportunities, the ongoing works, and results, identifying the challenges for future research.

\end{abstract}

\begin{IEEEkeywords}
LiFi technology, Visible Light Communications (VLC), 5G.
\end{IEEEkeywords}

\section{Introduction}
\label{sec:intro}

Nowadays, we are facing a deep electromagnetic spectrum 
crisis due to the explosive increase in wireless devices 
(currently, 16,000 million devices in the world, and with 
an expected increase of 23$\%$ for the year 2021).
The wide use of these devices is promoted by emerging 
multimedia applications such as augmented and virtual 
reality or video streaming services. These technologies require 
very high bandwidth and data rate, clogging the radio 
frequency (RF) band. A reference example is a pioneer and successful 
application \textit{Pokemon GO}, which throws 65 million of users connected 
to the Internet daily. On the other hand, even with advances in wireless technology, there are locations such as hospitals, aircraft or industries where the RF interferes with electronic devices, which can cause serious 
health issues.

In view of these problems, different signal-processing techniques 
-just as massive multiple-input multiple-output (m-MIMO) systems \cite{NextGeneration}-are considered as one of the key enabling technologies for future communications, such as 5G and beyond, to satisfy the expansion of network capacity.  
This is achieved due to the fact that they provide high spectral- 
and energy efficiency, which will be fulfilled by deploying many antennas in the base station (BS), far beyond those used in the current operational standards (Long Term Evolution, LTE, or 4G). However, the data rate is still not large enough to meet the demand of unforeseen applications. In addition, the issue of interference remains and increases with the use of many antennas. 

To circumvent all these impediments, the promising \textit{Light Fidelity} (LiFi) technology is emerging as one of the best technology of the Era 
to support the growing demand for higher transmission speeds as well as
ensuring RF interference-free environments.  
LiFi is based on the usage of visible light (VL) instead of RF in order to transmit information, offering 10,000 times larger bandwidth. Hence, using existing lighting infrastructure (streetlights, traffic lights, home or office lamps, flat panel displays, etc.) is proposed, which leads to other advantages. For example, achieving energy and cost savings which would change the cities towards more affordable and safer ones, known as \textit{Smart Cities}. 

Regarding lighting conditions and unlike infrared (IR), there are no health 
hazards of VL ($http://www.who.int/peh$-$emf/about/WhatisEMF$), it is thus applicable in the above-mentioned areas where RF-sensitive electronic devices are present. In addition, since the VL is confined by opaque walls, it provides a strong inherent security feature for LiFi over current WiFi (Wireless Fidelity).

Several studies on LiFi or Visible Light Communications (VLC) have taken into account the concept \cite{LiFi} and the technical aspects of the technology \cite{VLC_Tutorial}. Nevertheless, there is no reference that collects the key issues to design a LiFi system.  Neither a comparison of LiFi with other older wireless technologies for communications such as Bluetooth, ZigBee, Near Field Communication (NFC) or Worldwide Interoperability for Microwave Access (WiMAX) has been carried out regarding performance or scope. This study aims to explore the possible research areas, the current status, the ongoing projects, and the most important results for LiFi technology so far, as the trend will continue in coming years. Thus, this work is motivated by the need for a user-friendly research guide that enables a starting point for many future projects.

The novel contributions of this work are as follows:
\begin{itemize}
\item A current survey of the LiFi technology. This includes a state-of-the-art update of the 
technology and a comparison with other wireless technologies.

\item The first taxonomy is based on the different elements of a LiFi system.

\item Progress of activities and projects, just like the amazing opportunities which LiFi 
offers will be discussed, and a plan for the future will be drawn up. 

\item The latest improvements in the industry and academic world are compiled in this work.
\end{itemize}

The rest of the paper is organized as follows. In \refs{sec:over}
an overview of LiFi is presented. \refs{sec:comparacion} analyses the performance of LiFi versus other wireless technologies.
The taxonomy is proposed and discussed in detail in \refs{sec:taxonomy}. In \refs{sec:opport} the emerging opportunities are analyzed by highlighting their advantages over existing RF networks. The latest status of the results in research is shown in \refs{sec:result}. The challenges and the potential future of LiFi are discussed in \refs{sec:future}. 
Finally, \refs{sec:conclusion} presents the conclusions.

\section{Overview LiFi}
\label{sec:over}
This section reviews the concepts of a LiFi system. For the first time, the 
IEEE 802.15.7 Standard provides the properties for the Physical and  
Medium Access Control (MAC) layers of a VLC system. Later, Professor H. Haas was the first to suggest the concept of LiFi Network \cite{LiFi}.

\begin{figure}[ht]
\centering
\includegraphics[width=\linewidth]{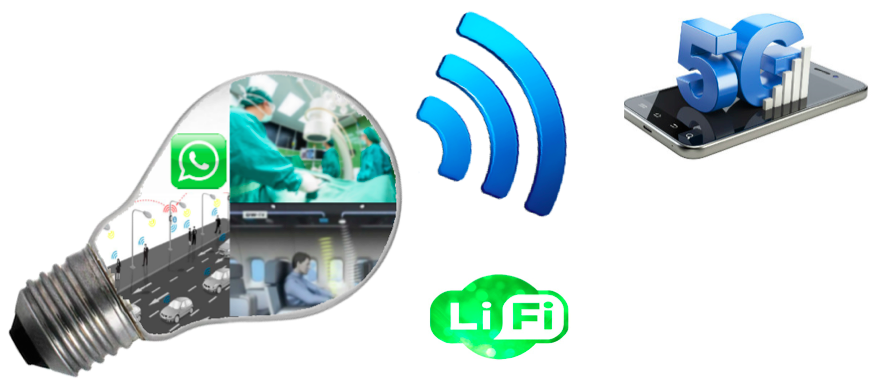}
\caption{Communication System for LiFi Technology.}
\label{f1_app}
\end{figure}

The basic components of the transmitter and receiver for a LiFi system are 
shown in \reff{f1_app}. 
The transmitter is a light source consisting of one or more LED (light-emitting diode) bulbs. The receiver is assumed 
by a photo-detector or an imaging sensor integrated into a mobile phone, just as in the case \reff{f1_app}, in a USB pen drive or a laptop.

The information is transmitted by the very high-speed changes of light intensity in the LED, which are unnoticeable by the human eye. This is equivalent to a digital signal, where the logical data '1' and '0' are encoded for being transmitted while the LED is ON and OFF, respectively. The LED flickers between 400 and 800 THz, corresponding to a nonlicensed visible light spectrum band which may provide enormous bandwidth. LiFi offers hundreds of THz against GHz provided by WiFi networks. This allows the optical carrier to transfer user data while illuminating at the same time.
Then, the light intensity is captured in the receiver. This element is the responsibility of introducing the limitations regarding high speed since it depends on the material it is made of. 
Examples of each element will be shown in \refs{sec:taxonomy} C. If we used the visible spectrum, we could achieve up to 100 Gbps as data rates. Still, more realistic implementations of this scheme have demonstrated that the LED can achieve transmitting data up to 10 Kbps and 500 Mbps using commercial normal fluorescent lamps. The maximum transmission distances attained have been at 1 and 2 kilometers for a low-rate data regime.

Regarding the communication channel, in \cite{Canal} was shown the channel model for LiFi. This is characterized by a line of signal (LOS) scheme, but it has been proved that it is not strictly necessary. Also, in a LiFi network, the channel would be affected by the non-linearity of the voltage and illuminance characteristics. Hence, we need to include schemes to dim the light. These are integrated into LiFi to save energy and facilitate intelligent lighting solutions with the aim of maintaining a constant brightness level that does not result in vision damage. The modulation scheme may help perform a dimming process inserted between the user data frames.

\section{LiFi versus Legacy Wireless Technologies}
\label{sec:comparacion}

The 5G is requiring new features and performance in future wireless communication systems. LiFi is a proposal in order to meet some requirements of the 5G. Therefore, a comparison of LiFi in terms of performance with respect to the legacy RF technologies that can be used for transferring data between devices nowadays is shown in \reft{table1}. 

The LiFi performance is mainly limited by interference from other light sources and obstacles which impact the coverage area. This is the major drawback of LiFi converting WiFi to its main competitor.  However, as we can see in \reft{table1}, LiFi offers a 10,000 times greater unlicensed spectrum by visible light waves than radio waves. Consequently, it is more attractive than WiFi for the future 5G.  For example, theoretically, using light as wireless internet access allows a high-definition film to be downloaded in 35 seconds, which is 250 times faster than broadband networks. These advantages outweigh those offered by fiber optic lines.
In addition, in the next year, thanks to its performance, LiFi could take over other technologies such as ZigBee, Bluetooth, or NFC. 
Although WiMAX or Long Term Evolution (LTE) has a big scope regarding the deployment, their present costs are very high compared to LiFi, also the antennas used have a high environmental impact.

The devices used are considerably cheaper than the RF counterparts, mainly WiFi, WiMAX, and 4G, which require a high cost of deployment. Regarding the rest, the cost is similar, but the materials used for LiFi show less aggression toward the environment.

The propagation losses are calculated for free space. LiFi still shows large losses for short distances against WiFi.

In terms of system complexity, EMI, power consumption, or latency, LiFi is similar to Bluetooth, ZigBee, or NFC. However, its performance makes it more competitive since it allows greater data density. In addition, unlike IR, visible light presents no health hazards in illumination conditions. Note also that LiFi owns an inherent security feature against RF systems, making transmission and reception simpler than in RF communication. The integration of LiFi is more accessible due to the fact that the lighting infrastructure is already exiting, which claims new laws to be enacted.

Based on this comparison and performance of LiFi, several opportunities, open issues, and future challenges emerge and will be detailed in \refs{sec:opport} and \refs{sec:future}.

\begin{table*}[]
\centering\caption{Comparison Between LiFi and Other Wireless Technologies: WiFI, WiMAX, LTE, Bluetooth, ZigBee and NFC.}
\label{table1} 
\begin{tabular}{c || c | c | c | c | c | c | c}
\hline
Feature &LiFi                   & WiFi   	        & WiMAX  & LTE  &      Bluetooh & ZigBee & NFC\\
             & IEEE 802.15.7 & IEEE 802.11g & IEEE 802.16e & (3GPP 4G)  & 5.0 & 802.15.4  &  \\
\hline
\hline
Frequency Range /  & 385-784 MHz & 2.4-5 GHz & 2-11 GHz & Multi band\footnote{LTE:} & 2.4-5 GHz & 868-868.8 MHz & 13.56 MHz \\
Available Spectrum & & & & & & 902-928 MHz & \\
                                & & & & & & 2402- 2482 MHz & \\
\hline
Bandwidth & Unlimited & 20 MHz & 20 MHz & 1,5,10 and 20 MHz & 1 MHz & 9.15 MHz & 13.56 MHz \\
\hline
Data Rate & 1 Gbps & 54 Mbps & 100 Mbps  & 86.5 Mbps & 1 Mbps  & 250 kbps & 424 kbps \\
\hline
Coverage & 10 meters & 100 meters & 80 Km & 290-950 m & 10 meters & 10 meters & 0.2 meters \\
                &                  & Widespread & Widespread & Widespread & Widespread & & \\
\hline
Propagation Losses & 170 dB & 86 dB & 140 dB & 125 dB (1 Km) & 2 dB & 3 dB & 0.5 dB \\
\hline
Power consumption & 1 mW & 100 mW & 30 W & 1.26 W & 1 mW & 1 mW & 1 mW \\
\hline
Latency Delay & Very Low & Medium & High & Very Low & Low & Low & Very Low \\
\hline
Safety & Unregulated & Intensity    & Intensity   & Intensity  & Intensity    & Intensity  & Intensity  \\
           &                      & Regulated & Regulated& Regulated & Regulated & Regulated & Regulated \\
\hline
Security & High & Limited & Limited & Medium & High & High & High \\
\hline
Cost & Low & Medium & High & High & Low & Low & Low \\
\hline
System Complexity & Low & High & High & High  & Low & Low  & Low \\
\hline
Environmental impact & None & Yes & Yes & Yes & None & None & None \\
\hline
Electromagnetic Interference & None & Yes & Yes & Yes & None & None & Yes \\
\bottomrule
\end{tabular}
\end{table*}

\section{Taxonomy}
\label{sec:taxonomy}
The LiFi technology opens the doors to numerous research areas. 
Therefore, we want to address for the first time a taxonomy that reflects all the 
current working points.

\begin{figure*}[]
\centering
\includegraphics[width=\linewidth]{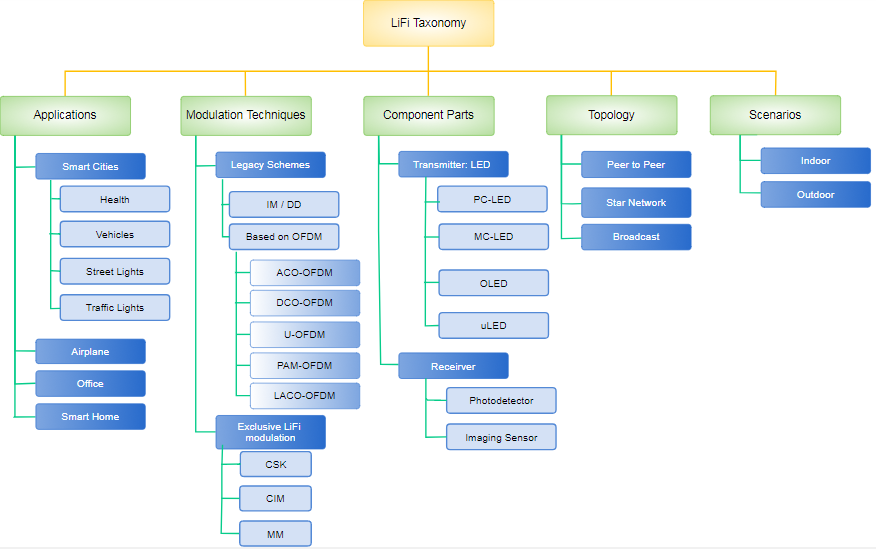}
\caption{Taxonomy for LiFi Technology.}
\label{f2_tax}
\end{figure*}

\reff{f2_tax} depicts the devised taxonomy, based  
on parameters such as applications, modulation techniques, 
the component parts and topologies of a LiFi system, as well as  
the possible scenarios.

\subsection{New Applications}

\begin{figure}[ht]
\centering
\includegraphics[width=\linewidth]{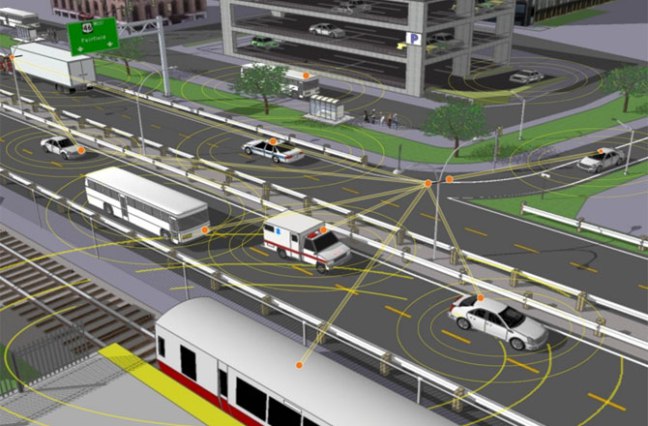}
\caption{Smart Cities using LiFi Technology.}
\label{f4_city}
\end{figure}

\begin{figure}[ht]
\centering
\includegraphics[width=\linewidth]{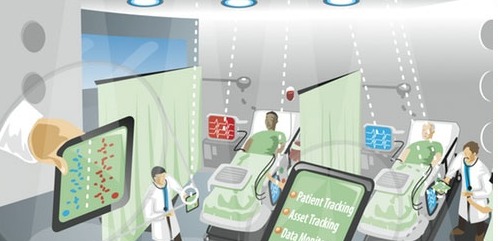}
\caption{Future Hospital using LiFi Technology.}
\label{f3_hosp}
\end{figure}

The absence of RF interference opens a wide range of new
applications which may be the initial point to designing a LiFi 
system. \textit{Smart Cities} are the main focus of technological 
innovation using LiFi \cite{SmartCity_App}. As shown in \reff{f4_city},
we can fully connect a city through light: road signs, public 
transportation systems, and ambulance services to hospitals, all of which 
increase the security level and quality of life of the population.
Several projects have been engaged in these apps. In 
\cite{VictorTestbed} is proposed the integration of LiFi in streetlights
to transmit the Internet by light. The healthcare system can benefit from LiFi by having access 
to the Internet, and for use with medical equipment, \cite{Hospital_App}. This can also be useful in robotic surgeries, avoiding the hazards of RF since it can penetrate through the human 
body, unlike light. In \reff{f3_hosp} can be seen a future hospital with LiFi integrated. 
In-home application, \cite{Home_App} shows a wide range of 
varieties suited to daily life in the home.
 
 In \cite{Automotive_App} showed great 
progress for the automotive sector.  Vehicle headlights are 
gradually being substituted with LEDs. This offers the hope 
of vehicle-to-vehicle communication (V2V-C) through LiFi.
Hence the communication between vehicles and the traffic 
lights is outstanding. In addition, V2V-C can offer the development 
of new and more powerful anti-collision systems to swap over 
information between vehicles while driving. This approach 
is also shown in \reff{f4_city}. Another emerging application is the Internet 
in Aircrafts \cite{Airplane_App} without obstructing the navigation 
systems. The reduction in cabling necessity also will lead to lighter 
aircraft. A new area is for education systems, where LiFi is the 
most recent technology that can give access to high-speed internet. Therefore, schools can use LiFi 
for video conferences, digital tutorial downloads, and online learning. 
Also, LiFi is a powerful solution for the chemical or petrochemical Industry 
where RF is greatly damaging.

\subsection{Modulation Techniques}

 An important factor is a way that we shape the light pulses.
Therefore, the modulation scheme is another design criterion.
Generally, classical \textit{intensity modulation and direct detection} (IM/DD) is employed for optical wireless communication: on-off keying (OOK), pulse position 
modulation (PPM) and pulse amplitude modulation (PAM), which may be easily implemented, but introduce a degradation in the performance as their spectral efficiency increases. Multiple carrier modulation schemes based on \textit{orthogonal frequency division multiplexing} (OFDM) are considered to be potential candidates
for optical wireless channels since they only require a single tap
equalizer at the receiver. In this case, different schemes are proposed:
the principle of \textit{asymmetrically clipped optical} (ACO-OFDM) is to only load 
the odd subcarriers with useful information. When a DC current is added, we have a
\textit{DC optical biased} (DCO)-OFDM scheme. A real unipolar optical OFDM is realized in PAM OFDM by exploiting the Fourier properties of imaginary signals. Another novel waveform proposed is \textit{unipolar} U-OFDM and \textit{layered asymmetrically clipped optical} (LACO) OFDM, which can be achieved by zero level clipping with the need of any DC bias.
 
On the other hand, new modulation techniques arise with LiFi. These use
the color dimension and add a new degree of freedom to transmit information. 
The idea is to change the colors which compose the white light. This is
easier than modulating the carrier frequency of the LED.
The IEEE 802.15.7 standard proposes \textit{color shift keying} (CSK) as
a modulation technique for VLC. The incoming bits are
mapped into a constellation of colors from the chromatic color space. 
Other techniques based on color dimension are: \textit{color intensity modulation} 
(CIM) or \textit{metameric modulation }(MM).
A more detailed explanation of the whole modulation pack for LiFi is shown in \cite{Modulation}.

\subsection{Component Parts}

Section II explains that the two basic elements for LiFi are a LED and
a photo-receptor (PD). We will achieve different data rates depending on 
the material that these components are made of. There are several LED 
categories currently available on the market that promote specific applications:

 \begin{itemize}
\item \textbf{PC-LEDs}: \textit{phosphorus compounds} for generating white light by varying the green, yellow, and red portion of the spectrum. 

\item \textbf{Multi-Chip (MC) LED}: set of three or more LED chips 
emitting different colors, typically red, green, and blue. Using multiple LEDs 
adds the benefits offered by m-MIMO systems \cite{NextGeneration}, eliminating their disadvantages.

\item \textbf{Organic Light Emitting Diodes (OLED)}: generate light using an 
organic layer unlike PC-LEDs or Multi-Chip, which makes OLEDs less suitable 
for high-speed applications. Their luminous efficiency has been improved at the 
 cost of a reduced lifetime.
\end{itemize}

PC-LEDs are cheaper and simpler than MC LEDs; however, they have a bandwidth limitation due to the low phosphor conversion efficiency. It is expected that the majority of the new energy-efficient lighting installations are expected to be based on LED technology by 2020, thus, prices of these devices will continue to fall in the coming years making this technology more attractive. 

On the receiver side, we have PDs and imaging sensors as possible candidates for detecting visible light. There are two types of photodetectors used in VLC systems: 

\begin{itemize} 
\item \textbf{PIN} (PD): has been used as predominant in VLC due to high-temperature tolerance and lower cost.

\item \textbf{Avalanche PD} (APD): are more expensive, although they may offer higher rates required for LiFi.
\end{itemize}
More technical details about optical components for LiFi in \cite{Video_Lajos}.

\subsection{Topology}
Different topologies are given depending on the number of LEDs available and the MAC protocols supported. Typical LiFi links use LOS configuration due to their illumination purpose. However, other configurations are possible such as star network or broadcast. In \cite{VictorTestbed} was shown a receiver LiFi with capabilities for signal detection by diffusion.
A recent research area regarding topology is to employ many LEDs as transmitters and receivers. This is equivalent to the m-MIMO system \cite{NextGeneration}. As mentioned above, an m-MIMO configuration improves the spectral efficiency beyond that of RF-based systems. On the other hand, the channel characteristics will influence the topology decision and the chosen modulation technique.

\subsection{Scenarios}
 
LiFi is more suitable for indoor scenarios such as hospitals, homes, or offices compared to the popular IR band, which is preferable for long backhaul communication. However, this is being raised for outdoor scenarios in the Smart Cities field, as shown in \cite{VictorTestbed}. 
Here, the main topic, such as the influence of external light (sunlight or artificial light), is considered. Thus, most of the projects are still in indoor scenarios. The choice of the scenario also influences the decision of topology and modulation techniques since all these features are linked through the channel between LED and PD.

\section{Opportunities}
\label{sec:opport}

The integration of LiFi in wireless devices and networks has brought new opportunities for future communication systems. The main ones are as follows:

\begin{enumerate}
\item \textbf{Very high data rate} \cite{Challenges}: higher bandwidth due to the more ample spectrum over the VL range guarantees the growing data demand by smartphones.

\item \textbf{Security}: LiFi signal exhibits confinement property, which deters hackers from accessing information from outside a building.  

\item \textbf{Free Electromagnetic Interference}: the use of VL instead of RF is safer in places where RF sensible devices are present (hospitals, airplanes, industrial zones).

\item \textbf{Spatial Reuse}:  LiFi signals in adjacent areas do not interfere with each other since light cannot penetrate through opaque objects. Therefore, we can reuse the optical carrier frequencies. 

\item \textbf{Energy Efficiency} (lower energy consumption): LEDs are energy efficient and highly controllable light sources, which allows them to be part of Green technology. Also, there is always some light source around us, facilitating the use of LiFI with minimum extra consumption.
 
\item \textbf{Low cost}: the implementation is cheaper than other communication systems since it is carried out in existing electrical Infrastructures.  
\end{enumerate}

\section{Results Status}
\label{sec:result}

Numerous companies and industries are being motivated by LiFi technology due to its potential in new business areas and sectors where WiFi has no place. This section discusses the currently open development projects Worldwide, which are marking a key turning point. Table I shows various companies' efforts to achieve a higher data rate using LiFi based on the new taxonomy presented in \refs{sec:taxonomy}.
The company \textbf{PureLiFi} was the first to develop a ceiling unit called \textit{Li-Flame} capable of transmitting 10 Mbps using a standard LED. The first LiFi solution with mobile coverage was presented at Barcelona's 2018 Mobile Word Congress (MWC). \textbf{OLEDCOMM} is a pioneer company in the introduction of LiFi with several projects\footnote{http://www.oledcomm.com/lifi-technology}, the most recent one a lamp with multiple LED, which was also presented at 2018's MWC. \textbf{Lucibel}, a French company, developed a full wireless network through a bidirectional link that reaches up to 42 Mbps\footnote{http://www.lucibel.com/lifi-haut-debit/}. 
In Mexico, \textbf{SISOFT} commercializes since 2016 the components needed for LiFi networks working to 200 Mbps. 
In Spain, \textbf{Indra Systems} developed a LiFi system that integrates audio-description information to bring more people into employment, focusing on particularly disadvantaged individuals.
Another project conducted in Spain was ``Smart Light" \cite{VictorTestbed}, where a proof of concept is aimed to incorporate the Internet into streetlights to develop the Internet of Things (IoT). Other main companies are \textbf{Velmenni} and \textbf{Stins Coman}, focusing on deployments of LAN LiFi networks.
\textbf{Philips} and \textbf{Acuity Brands} are performing tests in order to incorporate LiFi in indoor positioning systems, which aim for shopping centers or places with a high number of people. In addition, NASA predicts launching a satellite with LiFi which allows Geostationary satellites to communicate with Low-Earth Orbit satellites.
Microsoft is also implementing the LiFi solution at its innovation center in Issy-les-Moulineaux. Airbus develops a solution to incorporate LiFi into airplanes. Slux is a Swiss company that achieved transferred data up to 15 Mbps by VL without the need for LOS\footnote{https://www.slux.guru}.

\begin{figure*}[]
\centering
\includegraphics[width=\linewidth]{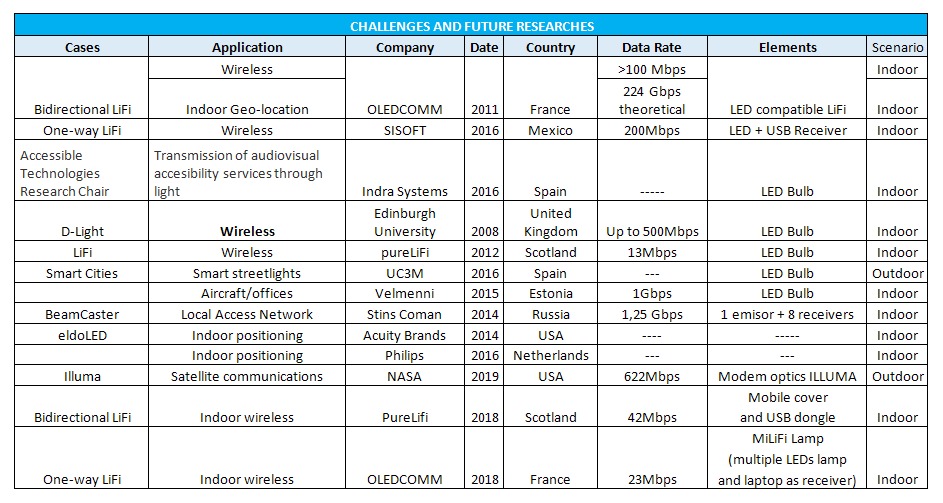}
\caption{Advances and Ongoing Projects for LiFi Technology.}
\label{f3_app}
\end{figure*} 

In the academic world, several groups of work are associated with LiFi or VLC:
\textbf{University of Edinburgh} was the first one aiming to develop a LiFi commercial solution. Its interest is in utilizing OFDM for high
speed data communication. The European \textbf{hOME Gigabit Access (OMEGA)} Project, funded by the European Commission and 19 partner institutions\footnote{Siemens AG, Fraunhofer Henrich-Hertz Institute, France Telecom, Infineon, the University of Oxford to mention but a few.},  achieved 100 Mb/s wireless connectivity to home users by VLC using OOK-NRZ modulation schemes and later with DMT modulation. After using an RGB LED and a PIN-type receiver, they attained a rate of 806 Mbps. \textbf{Oxford University} have been focused on employing equalization and MIMO techniques.  They demonstrated that using pre and post-equalizers in the transmitter and receiver improves the bandwidth by 3 dB and achieves a transmission up to 1.1 Gbps with pc-LEDs. \textbf{Keio University Group} includes system modeling using OFDM and OOK as modulation schemes. Their focus is intelligent transport systems utilizing existing traffic lights.
\textbf{Boston University} proposes RF/VLC hybrid methods that can avoid obstacles such as walls. \textbf{KAIST University} researches mainly dimming control techniques in VLC using modulation schemes.

\section{Remaining Challenges and Potential Future researches}
\label{sec:future}

LiFi systems, even though recent progress has been made, still face numerous open research issues which need to be addressed.  In addition, the current status of technology presented in this paper shows an idea that may be advocated as future lines in the research community in the years to come for this emerging technology.

\subsection{Open Research Issues}

Currently, the advances in LiFi are suffering some apparent restraint due to the nature of the technology. Hence, there are open research issues that must be solved. 
First, the speed rate is lower than LiFi may offer. Therefore, several techniques are being studied to improve the modulation bandwidth transmission devices used, e.g., blue filtering or other LED materials. Regarding the receiver, we need to include new control and synchronization schemes with the aim of reducing the influence of external light (natural or artificial). The current proposals in the literature show a reduced distance between transmitter and receiver, making them unfeasible for a real communication system; therefore, it is needed to include a scheme to boost the range and improve the coverage area. Another issue to resolve is to separate the signals from several light sources (users) on a single detector. This is necessary to ensure multiuser environments with LiFi. 

In the hardware Industry, companies are working towards the miniaturization of LiFi technology. Ultimately, the goal is to have LiFi on every mobile device. This means that the technology will be affordable for integration into handsets, tablets, and laptops. In addition, this technology needs to be integrated or replaced on the existing lighting infrastructures, turning them into \textit{Green Mobile-Lighting Communication Networks}. These networks will consume less energy and power.
Since LiFi does not work in your pocket, the roaming between technology, if the light signal comes from or to a LiFi-enabled device is below the receiver's threshold, it will not receive data. In that instance, radio systems or cellular networks, if available, will continue to deliver data. This type of network will be known as \textit{hybrid networks}. Another requirement is looking for an interface between LiFi devices since the present prototypes employ a base connected by USB 3.0. The problem is that this proposal can only offer the same data rate as Ethernet, not enough for future applications.

\subsection{Future Researches}

Mobile data traffic will reach the following milestones within the next 5 years. 
\begin{itemize}

\item Monthly global mobile data traffic will be 49 exabytes by 2021, and annual traffic will exceed half a zettabyte.

\item The average global mobile connection speed will surplus 20 Mbps by 2021, with more than 1.5 devices per capita at a time. There will be 11.6 billion mobile-connected devices by 2021, exceeding the world's projected population at that time (7.8 billion). 

\item 80\% of wireless data is consumed indoors, 70\% of that consumption will be video streams.

\end{itemize}

\begin{figure}[ht]
\centering
\includegraphics[width=\linewidth]{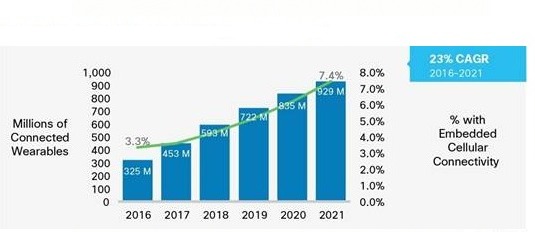}
\caption{Global Connected Wearables Devices at the year 2021$^1$.}
\label{f6_data}
\end{figure}

In view of these developments, new future lines emerge for LiFi technology during the next ten years: 

\begin{enumerate}

\item \textbf{Standardization of this technology}: at the moment, these tasks are carried out by Fraunhofer IPMS Company. In addition, LiFi is expected to take part in the coming Release of 3GPP for the 5G communication system. The implementation of this standard is expected in the early 2020s, merging the LiFi and 5G technologies.

\item \textbf{Spectrum Regulation}: due to the unlicensed visible spectrum or usage of exiting public infrastructures, legislation is necessary for the stability and proper development of the LiFi technology and to be placed on the market. This is the most important in the next 10 years.

\item \textbf{Uplink schemes}: all prototypes are considered downlink so far, but the increased traffic by 2021 is considered uplink too.

\item \textbf{Terminal mobility}: handover or changes between light sources are not considered yet beyond lab testing. 

\item \textbf{Coverage}: related to the before point, increasing the coverage area is a limiting factor to deploying a fully networking LiFi solution.

\item \textbf{Commercialization of LiFi} is likely going to come by way of an incremental strategy that first brings to market certain use cases of LiFi that are not unforeseen. There are certain businesses to introduce the communication feature in the illumination product since LiFi is capable of concurrently providing both features.

\item  \textbf{Unforeseen applications}: due to its directionality and containment properties, LiFi is also a good candidate for near-field communications (NFC) and a contender for providing indoor GPS capabilities. Various researchers are exploring light-based positioning and localization as potentially more accurate and easily deployed than RF or acoustic techniques. Recently, wearables and IoT are examples of applications that will claim large amounts of data as 
CISCO reveals\footnote{https://www.cisco.com/c/en/us/solutions/collateral/service-provider/visual-networking-index-vni/mobile-white-paper-c11-520862.html} and it is shown in \reff{f6_data} the evolution up to 2021. Therefore, LiFi has an imperative feature that connects everything to the Internet using LED lights as an Access Point in IoT.

\item \textbf{Spectrum Capability}: the spectrum now has to accommodate more mobile users. Connected devices are forecasted to increase to 20 billion IoT devices by the year 2021.  LiFi will use 6 THz for the next 10 years, 12,000 larger than WiFi. Hence, new multiplexing techniques will be needed.

\item \textbf{New waveforms}: New modulation schemes adapted to transmission by light in which the estimation of the channel is relieved by non-coherent schemes \cite{SingleRiceVictor,Thesis,Grouping}.

\item \textbf{Satellite Integration}: Incorporation of LiFi technology in satellite communications \cite{integration} as indicated by 3GPP, adapting new channel models \cite{channel}.

\end{enumerate}

In addition, in view of the performances provided, in 10 years LiFi will replace the ZigBee technology in home automation usage, Bluetooth, and NFC, merging them into only one technology.
 
\section{Conclusions}
\label{sec:conclusion}

This study performs an updated survey of the LiFi technology, which constitutes a foundation guideline for future research areas. LiFi is a powerful green technology that overcomes the drawbacks of the current RF-based technologies such as WiFi. The information transmission is by visible light. We have presented a taxonomy that collects the main current research areas and guidelines to design a LiFi system or new application. Furthermore, we analyze the most advanced projects so far, both in the academic and industry world. We highlight the opportunities LiFi offers against its counterpart WiFi. In addition, we propose challenges to be still addressed and the future lines with the expected date of this powerful technology for the next years.
 


\begin{thebibliography}{99}

\bibitem{NextGeneration} E.G. Larsson, O. Edfors, F. Tufvesson, and T.L. Marzetta,
        ``Massive MIMO for Next Generation Wireless Systems,''
        {\it  IEEE Comm. Mag.}, vol.~52, no.~2, pp.~186-195, Feb. 2014.
        
\bibitem{LiFi} H. Haas, L. Yin, Y. Wang and C. Chen, ``What is LiFi?,''
        {\it  Journal of Lightwave Technology.}, vol.~34, no.~6, pp.~1533-1544, Mar. 2016.
                
\bibitem{VLC_Tutorial} P.H. Pathak, X. Feng, P. Hu and P. Mohapatra  
``Visible Light Communication, Networking, and Sensing: A Survey, Potential and Chanllenges,''  
{\it  IEEE Communications Surveys \& Tutorials}, vol.~17, no.~4, pp.~2047-2077, 2015.

\bibitem{Canal}  M. Uysal, F. Miramirkhani, O. Narmanlioglu, T. Baykas and E. Panayirci, 
``IEEE 802.15.7r1 Reference Channel Models for Visible Light Communications,'' 
{\it IEEE Comm. Mag.} vol. 55, pp. 212-217, 2017.

\bibitem{SmartCity_App}  W. Boubakri, W. Abdallah and N. Boudriga 
``A light-based communication architecture for smart city applications,''  
{\it International Conference on Transparent Optical Networks (ICTON)} 2015, pp.1-6.

\bibitem{VictorTestbed} V. M. Baeza, M. Sánchez-Fernández, A. G. Armada and A. Royo, 
``Testbed for a LiFi system integrated in streetlights,''  
{\it 2015 European Conference on Networks and Communications (EuCNC)}, Paris, France, 2015, pp. 517-521.

\bibitem{Hospital_App} C.A. Shivakumar and P. Rajeshwari   
``LiFi Based Advanced Patient Monitoring System,''
        {\it  International Journal of Engineering Science and Computing.}  pp.~8212-8215, Jul. 2016.

\bibitem{Home_App}   C. W. Chow, Y. Liu, C. H. Yeh, J. Y. Sung and Y. L. Liu, 
``A practical in-home illumination consideration to reduce data rate fluctuation in visible light communication,'' 
{\it IEEE Wireless Comm.} vol. 22, pp. 17-23, 2015.


\bibitem{Automotive_App} A-M. Cailean and M. Dimian 
``Impact of IEEE 802.15.7 Standard on Visible Light Communications Usage in Automotive Applications,''  
{\it IEEE Comm. Mag.} pp. 169-175, Apr. 2017
   
	
\bibitem{Airplane_App} C. Quintana, V. Guerra, J. Rufo, J. Rabadan and R. Perez-Jimenez
``Reading lamp-based visible light communication system for in-flight entertainment,''
{\it IEEE Transactions on Consumer Electronics} vol. 59, pp. 31-37, 2013.


\bibitem{Modulation} M. S. Islim and H. Haas ``Modulation Techniques for LiFi,'' 
                     {\it ZTE Communications} vol. 1, no. 2, Apr. 2016.

\bibitem{Video_Lajos}  J. Jiang {\it et al}. ``Video streaming in the multiuser indoor visible light downlink,''
{\it IEEE Access}, vol. 3, pp. 2959-2986, 2015.


\bibitem{Office_App} O. Bouchet {\it et al} ``Visible-light communication system enabling 73 Mb/s data streaming,''
{\it IEEE Globecom Workshops} pp. 1042-1046, 2010.


\bibitem{Experimental} A. Burton, H. Le Minh, Z. Ghassemlooy, E. Bentley and C. Botella,
``Experimental Demonstration of 50-Mb/s Visible Light Communications Using 4x4 MIMO,''  
 {\it IEEE Photonics Technology Letters} vol. 26, pp. 212-217, 2014.
  
  
\bibitem{Challenges} D. Tsonev, S. Videv and H. Haas,  
``Towards a 100 Gb/s visible light wireless access network,'' 
{\it Optics Express}, vol. 23, no. 2, pp. 1627, 2015.

\bibitem{SingleRiceVictor} V. M. Baeza and A. G. Armada, ``Analysis of the performance of a non-coherent large scale SIMO system based on M-DPSK under Rician fading,'' \textit{2017 25th European Signal Processing Conference (EUSIPCO)}, Kos, Greece, 2017, pp. 618-622, doi: 10.23919/EUSIPCO.2017.8081281.

\bibitem{Thesis} V.M. Baeza, \textit{Multiuser non coherent massive mimo schemes based on dpsk for future communication systems} (Doctoral dissertation, Universidad Carlos III de Madrid) 2019.

 \bibitem{Grouping} V. M. Baeza and A. G. Armada, ``User Grouping for Non-Coherent DPSK Massive SIMO with Heterogeneous Propagation Conditions,'' \textit{2021 Global Congress on Electrical Engineering (GC-ElecEng)}, Valencia, Spain, 2021, pp. 26-30, doi: 10.1109/GC-ElecEng52322.2021.9788350.

\bibitem{integration} Monzon, V., Ha, V. N., Querol, J., $\&$ Chatzinotas, S. (2022). Non-Coherent Massive MIMO Integration in Satellite Communication. ArXiv. https://doi.org/10.48550/arXiv.2209.09701

 \bibitem{channel} V. M. Baeza, E. Lagunas, H. Al-Hraishawi and S. Chatzinotas, "An Overview of Channel Models for NGSO Satellites," 2022 IEEE 96th Vehicular Technology Conference (VTC2022-Fall), London, United Kingdom, 2022, pp. 1-6, doi: 10.1109/VTC2022-Fall57202.2022.10012693.
 
\end{thebibliography}
\end{document}